\documentclass[conference]{IEEEtran}
\usepackage{cite}
\usepackage{amsmath,amssymb,amsfonts}
\usepackage{algorithmic}
\usepackage{graphicx}
\usepackage{textcomp}
\usepackage{xcolor}
\usepackage{tabularx}
\usepackage{multirow}
\usepackage{multicol}
\def\BibTeX{{\rm B\kern-.05em{\sc i\kern-.025em b}\kern-.08em
    T\kern-.1667em\lower.7ex\hbox{E}\kern-.125emX}}
\usepackage{hyperref}
\begin{document}

\title{On the Analysis of Computational Delays in Reinforcement Learning-based Rate Adaptation Algorithms\\
}

\author{
    \IEEEauthorblockN{
        Ricardo Trancoso,
        Ruben Queiros,
        Helder Fontes,
        Rui Campos
    }
    \IEEEauthorblockA{
        INESC TEC and Faculdade de Engenharia, Universidade do Porto, Porto, Portugal \\
        \{ricardo.j.espirito, ruben.m.queiros, helder.m.fontes, rui.l.campos\}@inesctec.pt
    }
}

\maketitle

\begin{abstract}

Several research works have applied Reinforcement Learning (RL) algorithms to solve the Rate Adaptation (RA) problem in Wi-Fi networks.
The dynamic nature of the radio link requires the algorithms to be responsive to changes in link quality.
Delays in the execution of the algorithm may be detrimental to its performance, which in turn may decrease network performance.
This aspect has been overlooked in the state of the art.

In this paper, we present an analysis of common computational delays in RL-based RA algorithms, and propose a methodology that may be applied to reduce these computational delays and increase the efficiency of this type of algorithms.
We apply the proposed methodology to an existing RL-based RA algorithm.
The obtained experimental results indicate a reduction of one order of magnitude in the execution time of the algorithm, improving its responsiveness to link quality changes.

\end{abstract}

\begin{IEEEkeywords}
Reinforcement Learning, Rate Adaptation, Computational Delay
\end{IEEEkeywords}

\section{Introduction} \label{s1}

The IEEE 802.11 standard, commonly known as Wi-Fi, enables the establishment of Wireless Local Area Networks.
The standard has had many amendments to keep it up-to-date with the growing requirements of the vast application scenarios.
For example, new configuration parameters, such as more spatial streams and larger channel bandwidth, were introduced in more recent variants \cite{axStandard}. 
When configured correctly for a given link quality, these parameters enable increasing the efficiency of Wi-Fi networks.
One of the parameters that can be changed is the Modulation and Coding Scheme (MCS).
Since the link conditions are not static, the MCS has to be changed dynamically.
This is called Rate Adaptation (RA).

There are multiple heuristic-based algorithms for dynamic RA.
The most commonly used in practice are Minstrel \cite{minstrelsite} and Iwlwifi \cite{Grunblatt_2019}.
However, these algorithms have shortcomings \cite{chenExperienceDrivenDesign2021}, which limit their ability to select the optimal MCS rate.
Algorithms that employ Reinforcement Learning (RL) techniques have emerged as an alternative \cite{chenExperienceDrivenDesign2021,Karmakar_2017,karmakarIEEE80211ac2017,karmakarOnlineLearningApproach2020,pesericoRateAdaptationReinforcement2020,Karmakar2016}.
RL techniques collect observations from the environment.
With the observation, the algorithm takes a decision on the best action.
Finally, the algorithm calculates a numerical reward, which indicates how suitable the action was.
The algorithm repeats actions that yielded a high reward and avoids actions which led to the opposite.
Through this process, the algorithm can eventually find solutions to a problem automatically and even adapt to unforeseen circumstances.
In the context of RA, the environment is the radio link, and the algorithm learns how to configure the Wi-Fi parameters to increase efficiency in dynamic link conditions.
Since link conditions are constantly changing, an RL algorithm requires up-to-date observations and actions.
Therefore, delays during the execution of an RL-based RA algorithm may be detrimental to its performance.

In the state of the art, computational delays of RL-based RA algorithms have been overlooked.
Usually, only the conceptual stages of the algorithms are described, such as the RL model used and the information collected for the observations and
 the reward; other implementational details are typically not referred.
In \cite{chenExperienceDrivenDesign2021}, the authors mention the use of an asynchronous framework to prevent halting of the algorithm while waiting for a process to finish, but do not provide more details.
In \cite{Queiros_2022,Karmakar_2017,karmakarIEEE80211ac2017,karmakarOnlineLearningApproach2020,pesericoRateAdaptationReinforcement2020,Karmakar2016}, computational delays are not even mentioned.
Characterising the computational delays, which are not modelled in simulation, is relevant since they may impact the algorithm's real-world performance.

The main contributions of this paper are three-fold.
First, we bring up awareness to the execution time problem and the importance of considering computational delays when it comes to RL-based RA algorithms; this has been overlooked in the state of the art and may reveal that several existing algorithms are invalid.
Second, we propose an analysis methodology for RL-based RA algorithms, which separates implementational details from their conceptual design.
This can guide implementation alternatives that may be used to reduce the execution time and make RL-based RA algorithms valid in practice.
Finally, we apply the proposed analysis methodology to an existing algorithm and show the impact of different implementation alternatives in terms of execution time.

The paper is structured as follows.
\hyperref[s2]{Section~2} characterises the problem. \hyperref[s3]{Section~3} discusses our proposed implementation methodology, from a general overview to a more detailed description of some of the alternatives we evaluated.
\hyperref[s4]{Section~4} presents the experimental results we obtained by comparing those alternatives.
Finally, in \hyperref[s5]{Section~5}, we draw the conclusions.

\begin{figure*}[t]
\centerline{\includegraphics[width=\textwidth]{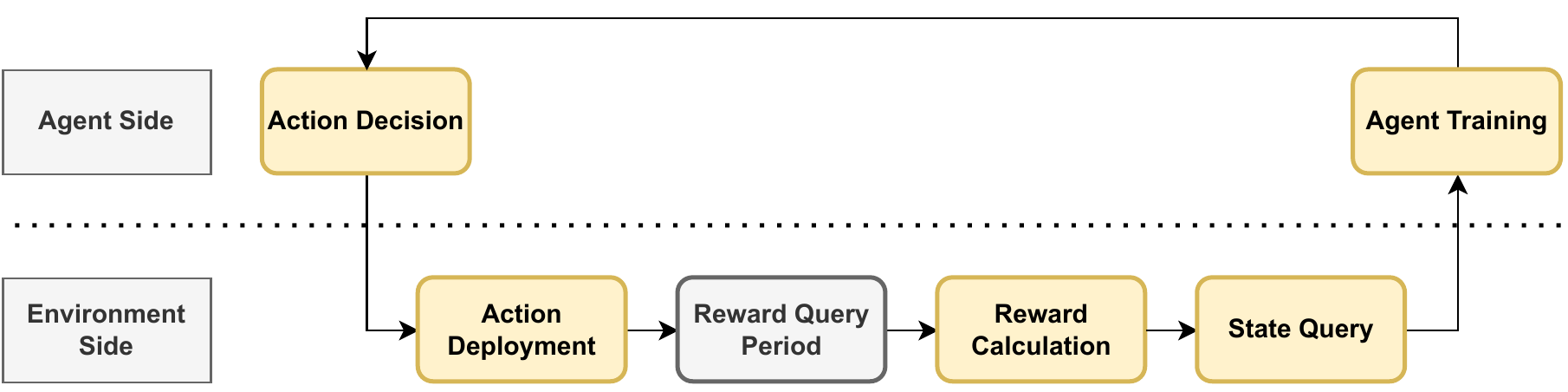}}
\caption{Overview of the DARA algorithm with its main operations highlighted, and split between agent and environment side.}
\label{fig:edaraoverview}
\end{figure*}

\section{Problem Characterisation}  \label{s2}

In this section, we characterise the problem by using the Data-Driven Algorithm for Rate Adaptation (DARA) \cite{Queiros_2022} as a representative state of the art RL-based RA algorithm for Wi-Fi networks.
In \cite{Queiros_2022}, DARA was evaluated in simulation, without considering any computational delays.
In order to characterise the problem, we first implemented DARA in Linux for allowing experimentation in a real wireless device.
This preliminary implementation used a simple user-space approach that neglects the computational delays incurred, in order to understand their impact in the performance of an RL-based RA algorithm.

In its original version, DARA uses a Deep Q-Network (DQN) \cite{mnih2013playing} agent that decides an action out of the eight possible MCS rates defined in IEEE 802.11n, considering a 20~MHz channel bandwidth and 800~ns guard interval.
Those rates have a theoretical maximum throughput of 6.5, 13, 19.5, 26, 39, 52, 58.5 and 65~Mbit/s.
In \hyperref[fig:edaraoverview]{Fig.~\ref*{fig:edaraoverview}}, we can see an overview of the DARA operations, based on the well known RL loop.
DARA begins with a state, which is the Received Signal Strength (RSS), in dBm, and decides its next action with this information, thus initiating the loop.
Then, it deploys that action and queries the reward statistics by collecting information on frame transmission successes and attempts.
The reward is calculated using Eq.~\ref*{eq}.
\begin{equation}
\text{Reward} = \text{FSR} \times \frac{\text{Current Rate}}{\text{Max Rate}}
\label{eq}
\end{equation}
This equation relates the frame success rate (FSR) of the current MCS with the theoretical throughput that it can achieve (Current Rate), in comparison to the maximum theoretical throughput of the highest MCS (Max Rate).
Thus, the reward will range between zero and one.
A reward of one implies the highest rate is being transmitted without errors.
It is important to note that while querying the reward, DARA waits a configurable amount of time so that statistics can be collected using the new MCS.
We call this time the \textit{reward query period}.
Afterwards, the algorithm queries the RSS of the device once again, which will be the state of the next action.
Finally, the agent trains its neural network model based on the reward that was received, and the loop repeats.

Because our implementation is in Linux, we attempted a simple user-space approach that uses bash commands for these operations.
These commands obtain information on the quality of the connection and the reward of the agent's actions.
In the state query, in order to measure the link quality, we read a file that contains the RSS.
To calculate the reward, we read a different file that contains the number of successful frame transmissions and total attempts.
This file exists in the Linux system by default, since it is generated by Minstrel, the default RA algorithm.
However, this file is only updated once every 100~ms, which is Minstrel's own ``reward'' query period.
This wait time, along with the time it takes to access the data on the device or process the commands, represents some of the existing computational delays present in experimental and real environments.
In simulation, these computational delays are neglected.

After testing one thousand execution steps, we measured experimentally that the response time of our simple approach was 599.4~ms on average.
This is a substantial increase in response time in comparison to the simulation results of DARA.
This increase is due to computational delays that are present in experimental environments, but are not modelled in simulation.
This demonstrates that disregarding these delays may hinder an algorithm's responsiveness.
For this reason, herein we propose an analysis methodology with the goal of characterising and reducing these delays, exploring alternative implementation strategies while maintaining the algorithm's original design.

\section{Analysis Methodology and Alternative Implementations} \label{s3}

After performing a solution-agnostic analysis of RL-based RA algorithms, we concluded about the key operations that these algorithms usually share.
We identified five main operations: 1) \textbf{action decision}, using the agent to choose the optimal expected action; 2) \textbf{agent training}, using the reward so that the agent may learn from its previous actions; 3) \textbf{action deployment}, applying the chosen action; 4) \textbf{reward calculation}, gathering information and calculating the numerical value of the reward; and 5) \textbf{state query}, gathering data on the current state of the environment.
By separating the implementation details of the algorithm from its design, different implementations may reduce the computational delays.

Given that these five main operations, common to state of the art algorithms, are present in DARA, hereafter we consider it as an application example of our analysis methodology, and propose different and comparable implementation alternatives.
In \hyperref[fig:edaraoverview]{Fig.~\ref*{fig:edaraoverview}}, these five operations are highlighted in orange and are split in two sides of the algorithm: the agent side; and the environment side.
The agent side comprises the action decision and the training of the algorithm.
It is more closely tied to the design of the algorithm since it is related to the mathematical techniques that RL approaches employ.
The environment side is responsible for interfacing with the device to deploy the action, calculate the reward, and query the state.

In what follows, we describe three different implementation alternatives of DARA that are based on the previously mentioned five key operations, while maintaining DARA's design.
Unlike the simple approach, these alternatives apply kernel-space techniques by modifying the mac80211 Linux kernel module.
These implementation alternatives are focused on the environment side because the agent side is tied to the design of the algorithm.

Additionally, we go over different agent implementations to measure whether changes to this side also incur different computational delays.
Because these agent side changes do alter the design of the algorithm, this was done from an exploratory perspective, without evaluating their impact in the performance of the algorithm.

\subsection{Environment Side Implementation and Information Parsing} \label{3b}

In the course of the reward and state query, the algorithm needs to gather information.
This information is the RSS, in the case of the state query, and statistics related to frame transmissions when it comes to the reward calculation.
This information comes from files in the device that need to be read and processed.
The files often contain information beyond the information needed, or they are read based on a non-ideal data type (e.g., as a string rather than as an integer).
For this reason, in addition to reading the files, it is also important to parse them in order to extract the required data.
Because this is a crucial step in the algorithm's functioning, it is worthwhile to explore efficient ways for both collecting the information and parsing it.

We considered three alternative implementations to collect the information:
\begin{itemize}
    \item \textbf{Subprocess} -- A Python module to spawn and interact with Linux shells, which enables the use of bash commands. This option was used in the simple DARA implementation.
    \item \textbf{Pure Python} -- A pure Python approach that uses the built-in \textit{read} function in order to access the files. This method may avoid the overheads of the previous option.
    \item \textbf{Rust extension} -- A Rust-based approach. Rust is a compiled language. It is expected to be faster than Python. This option attempts to see if using a compiled language to extend Python, without replacing it, is beneficial.
\end{itemize}

As for parsing the information, we evaluated three alternative implementations:
\begin{itemize}
    \item \textbf{Subprocess} -- It enables the use of bash commands and it can pipe the output through multiple filtering commands. This option was used in the simple DARA implementation to serve as a baseline for comparison.
    \item \textbf{Pure Python} -- We considered the use of Python built-in string functions on the output to filter it.
    \item \textbf{Regex} -- We considered the use of Regex search patterns to find and extract the information required from the output.
\end{itemize}

These alternatives are not an exhaustive list of possible implementations.
Our goal is to demonstrate how implementation differences affect computational delays, and compare the alternatives.

\subsection{Changes to the mac80211 Kernel Module} \label{3c}

The file read by the algorithm to obtain the RSS during state query is present on a Linux Ubuntu 14.04 distribution.
However, the file containing statistics on frame transmissions is not.
The simple DARA implementation obtained this information through Minstrel, a heuristic RA algorithm implemented in Linux.
Minstrel is implemented using the \textit{mac80211} Linux kernel module.
This module is used for managing wireless devices, thus having access to low-level information on wireless connections.
Nevertheless, this low-level information is usually locked to the kernel, and cannot be readily accessed, which is why the simple implementation reads from a table that Minstrel computes every 100~ms.
This table, unlike the data in the kernel module, is accessible in user-space, although the data is not as up-to-date due to the 100~ms period.
This step alone forces the first implementation to take at least 100~ms on each action decision.

For these reasons, our newer implementations of DARA modify the mac80211 module to create a virtual empty file that when read, runs a function that returns the values stored in the variables related to frame transmission successes and attempts.
We obtain accurate and current information on frame statistics, without being limited by the 100~ms waiting period associated with Minstrel's table update.
However, we still configure a \textit{reward query period} before calculating the reward, so our chosen action can take effect on the wireless connection.
This period is fully configurable, but has a tradeoff involved.
A smaller period means the algorithm can learn and take actions more frequently, but it may affect the learning performance of the algorithm if the collected data is not enough to be statistically relevant.
On the other hand, a longer period may improve the accuracy of the statistics, but may also miss the optimal time to switch rates.
We decided to use a 50~ms \textit{reward query period}, which still resulted in a total execution time below 100~ms.
This configurable period can be tuned further depending on the dynamics of the scenario, if a more rigorous analysis of the algorithm's performance is considered.

\subsection{Agent Implementation Details and Tradeoffs} \label{3d}

While the previous changes were part of the environment side of the algorithm, we briefly considered two questions related to the agent side.
We wanted to know how the computational delays of the algorithm are impacted by training alongside deployment and by using a different RL agent type rather than a DQN.
Although this is a conceptual difference that changes how the algorithm performs, this was done from an exploratory perspective.
In \hyperref[4d]{Section~4.D} we show that the difference in computational delays is not negligible.

RL has the capability of online learning.
This is when an algorithm is deployed and is performing its task, but it is still considering reward values from the actions it takes.
This enables the algorithm to keep learning, being able to adapt to unforeseen circumstances during its training phase.
To acquire information on computational delays, we tested the processing time of DARA in two situations.
First, considering it is constantly training at each new step.
Then, considering a situation in which it has trained beforehand and thus skips the training phase in deployment (i.e., online against offline learning).

As for the alternative to a DQN, we considered the use of Q-learning \cite{watkins1992q}.
Instead of modelling a state-action function through a neural network like a DQN, Q-learning stores a table of state-action pairs.
This table needs to be updated when training, which can be computationally expensive for large tables.
This is not the case for DARA, as it only considers eight rates, and discrete states, which is why we explored the Q-learning method.
We measured the action decision and training times of a Q-learning agent and compared them to the execution times obtained for a DQN.
Because these experiments were done only to evaluate the possible differences in computational delay related to the type of RL agent used, the actual network performance achieved by each RL algorithm was left for future work.

\section{Experimental Results} \label{s4}
\subsection{Methodology} \label{4a}

The DARA algorithm was implemented mainly in Python.
For this reason, we separated different operations of the RL algorithm, such as the action deployment or the state query, in different Python functions.
This enabled us to measure the time each function takes to run, which translates to the delay introduced by each operation.
The \textit{timeit} module \cite{timeit} was used for this purpose.

All tests were done in an experimental setting.
We used the w-iLab.2 testbed \cite{wilab} provided by the Fed4FIRE+ project.
Our network topology consisted of two nodes: a ZOTAC node as the access point, and a DSS node as the client.
The DARA algorithm was ran on the client node.
The client node was equipped with an Intel Core i5 2.6~GHz 4-core CPU, a 60~GB SSD hard drive, and a 4~GB DDR2 800~MHz PC2-6400 CL6 RAM.
The device ran a Ubuntu 14.04 Linux distribution, on kernel version 3.13.0.

There were multiple delays we analysed and attempted to minimise, such as the delay associated with parsing information and accessing a file.
For each of these delays, we considered different possible implementations, as detailed in the previous section.
We measured the time the alternatives take and determined the one that takes the lowest time.
In the end, we make decisions on which alternatives shall be used in each case.
This final version of DARA was then compared to the simple DARA implementation referred in \hyperref[s2]{Section~2}.

\subsection{Information Parsing} \label{4b}

There are two scenarios where information parsing is required.
These are: 1) during the state query; 2) during the calculation of the reward.
We will refer to them as \textit{state} and \textit{reward} scenario, respectively.
The \textit{state} file has a complex structure, with a large amount of information besides the value we want to extract, while the \textit{reward} file is the virtual file originating from our changes to the mac80211 module.
It reads variables directly from the module and only outputs two values separated by a comma, thus having a simpler structure.
For each scenario, we evaluated the three aforementioned alternatives:
1) using shell commands and shell command piping through the use of the subprocess module;
2) using a pure Python approach through string functions;
and 3) using Regex search patterns.

We measured ten thousand times the time each approach took to filter the data.
This measurement was repeated five times, and only the fastest measurement was considered.
Slower speeds are very often caused by other processes interfering with timing accuracy, and not due to variability in Python's speed \cite{timeit}, so this reduces the effect of these delays.
Finally, we calculate the average.
The time to load the required modules and one-time initialisations like compiling the regex pattern are not included.

\begin{table}[tbp]
    \caption{Average time of each information parsing alternative}
    \label{tab:infoparsingres}
    \begin{center}
    \begin{tabular}{ | c || c | c | c | }
    \hline 
    Scenario & Subprocess & Python & Regex \\ 
   \hline \hline
    State average (ms) & 5.0318 & 0.0017 & 0.0014\\  
   \hline 
    Reward average (ms) & 9.9792 & 0.0012  & 0.0018\\
    \hline
    \end{tabular}
    \end{center}
\end{table}

In \hyperref[tab:infoparsingres]{Table~\ref*{tab:infoparsingres}}, we can see the average time of one call to each of the information parsing functions.
The Python and Regex approaches are faster than the subprocess approach. 
Using them instead of subprocess can save around 15~ms for each algorithm step.
For the state scenario, Regex was slightly faster, while for the reward scenario, Python was the fastest method.

These results are related to the complexity of the files.
Because the state file is more complex than the reward file, it benefits more from the compiled search patterns of Regex.
However, the difference is small and either approach is reasonable.
In the end, we decided to use the Python parsing method for the file with reward statistics, and the Regex method for the file with state information as they are the fastest for each scenario.

\subsection{Environment Side Comparison} \label{4c}

As previously stated in \hyperref[3b]{Section~3.B}, there were three different implementations of the environment side, which differed in how they accessed files on the device.
Those were using the subprocess module, reading files through Python, and running a Rust extension to read the files.

To compare the three alternatives, we initialised the algorithm and ran each alternative for ten thousand steps.
The individual time of each information collection operation (i.e., state query, reward calculation) was measured, for each approach.
Afterwards, we calculated the average.
The measurements were done by recording the clock time before and after each operation, and calculating their difference.
To ensure we were only evaluating these three alternatives, all other algorithm parameters were kept the same.

\begin{table}[tbp]
\caption{Environment step average times per environment approach}
\begin{center}
    \begin{tabular}{|c||c|c|c|}
        \hline
        Environment & Subprocess & Python & Rust  \\
        \hline \hline
        step average (ms) & 99.637 & 62.805 & 65.107 \\
        \hline
        get\_reward average (ms) & 13.226 & 0.246 & 0.085 \\
        \hline
        get\_state average (ms) & 13.468 & 0.299 & 1.012 \\
        \hline
        set\_action average (ms) & 11.535 & 15.105 & 14.981 \\
        \hline
    \end{tabular}
\end{center}
\label{tab:envcomp}
\end{table}
    
\begin{figure}[tbp]
    \centerline{\includegraphics[width=\linewidth,height=5.5cm]{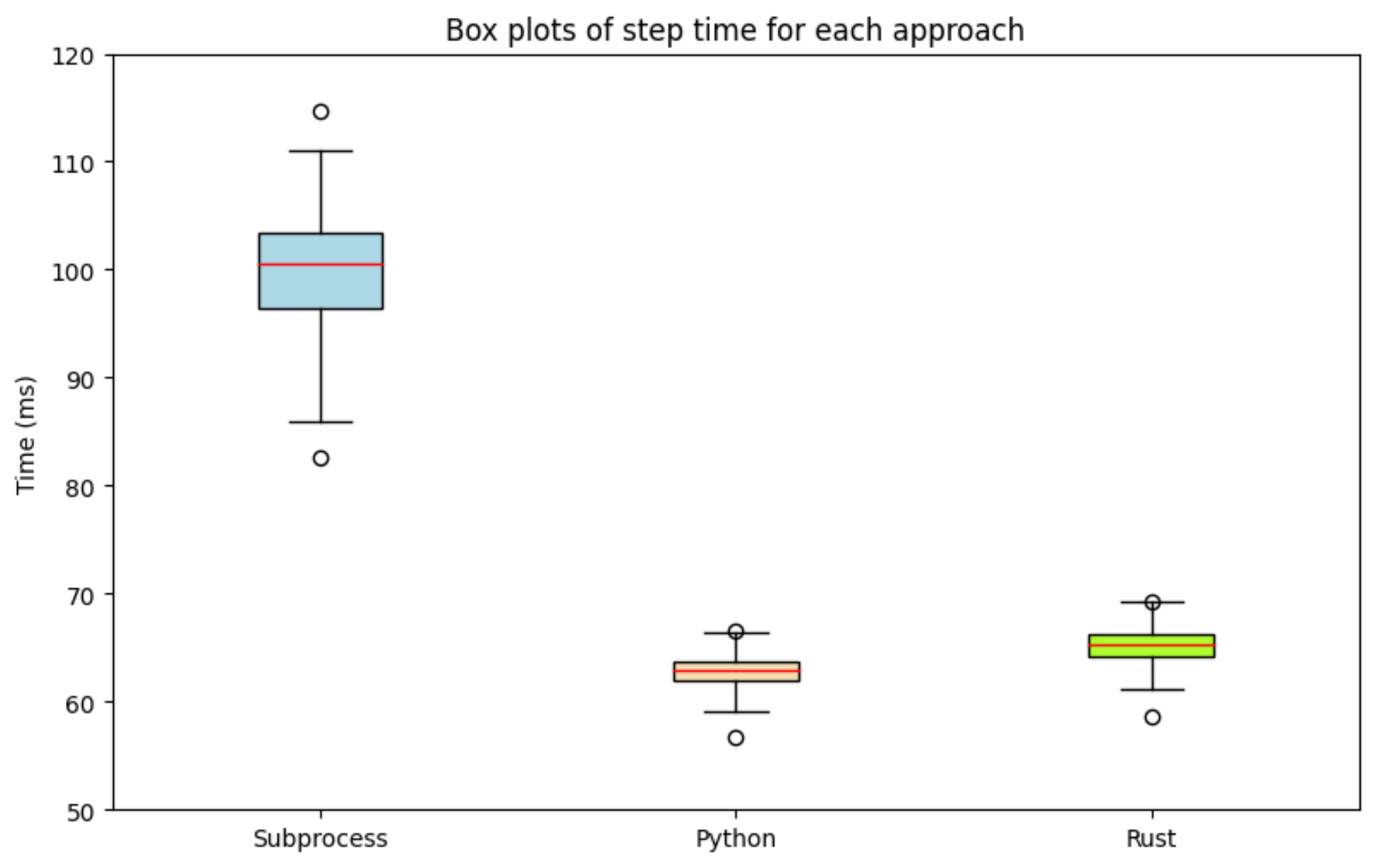}}
    \caption{Box plots of the environment step time for each environment side approach.}
    \label{fig:envcomp}
\end{figure}

In \hyperref[fig:envcomp]{Fig.~\ref*{fig:envcomp}}, we have a box plot for each approach that represents the time of an environment side step for that approach.
The approaches are represented on the x-axis, while the y-axis represents the time in milliseconds that each approach takes for each algorithm execution step.
The y-axis starts at 50~ms because that is our predetermined \textit{reward query period}.
The lower and upper boundary of the boxes represent the first and third quartiles, respectively.
The lower and upper whiskers represent the minimum and maximum recorded value within 1.5 times the interquartile range below or above the first and third quartiles, respectively.
The dots represent the minimum and maximum recorded values.
\hyperref[tab:envcomp]{Table~\ref*{tab:envcomp}} contains information on how much of that time is spent, on average, on the two operations our approaches modify, which are the state query and the reward calculation.
The time spent deploying the action is also represented for the sake of completeness, but between all three approaches, there were no differences in how it was implemented.
The action is deployed through a Linux terminal utility called \textit{iwconfig}.
Further changes to the Linux kernel may improve this operation, as it was the case for the reward query.
Despite the similar implementation, the action deployment appears to have been faster in the subprocess approach.

Overall, though, the subprocess approach is the slowest, with the other two taking 45~ms less each step.
The Python and Rust approaches are close, but Python is slightly faster.
The Rust approach in our implementation is an extension to the Python base, in an attempt to leverage Rust's faster execution speed.
However, for Rust function calls there is an additional overhead when converting a Python variable to a Rust variable, which may explain why it is slightly slower.
For this reason, we decided to use the Python approach in our final implementation of DARA.

\subsection{Agent Side Comparison} \label{4d}

In \hyperref[3d]{Section~3.D}, we considered two questions: 1) what difference does training make in terms of processing time; 2) what difference does another RL approach such as Q-learning make.
For this reason, we ran a similar test to the environment side comparison but we instead compared three different versions of the algorithm's agent.
Those versions were: a DQN agent with online learning enabled; a previously trained DQN agent that does not train during deployment; and a Q-learning agent with online learning enabled.
For each approach, the algorithm is initialised and ran for ten thousand steps.
The time the algorithm spends deciding the action and training is measured and averaged afterwards.

\begin{table}
\caption{Environment step average times per agent approach}
        \centering
        \begin{tabular}{|c||c|}
            \hline
            Agent & Average time (ms)  \\
            \hline \hline
            DQN (training) & 21.606 \\
            \hline
            DQN (no training) & 6.915 \\
            \hline
            Q-learning (training) & 3.062 \\
            \hline
        \end{tabular}
    \label{tab:agentcomp}
\end{table}

\begin{figure}[tbp]
    \centerline{\includegraphics[width=\linewidth,height=5.5cm]{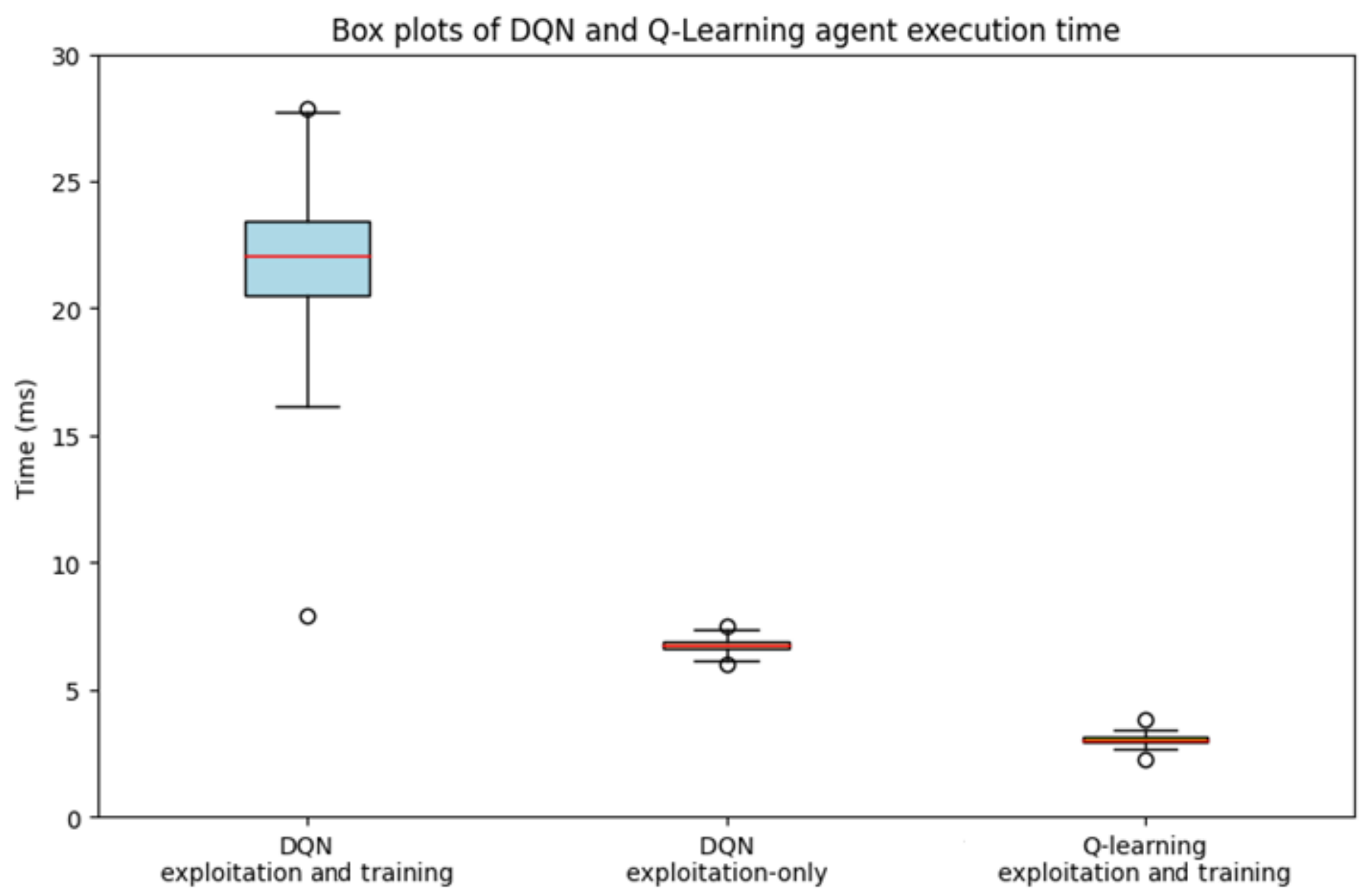}}
    \caption{Box plots of an environment step time for each agent approach.}
    \label{fig:agentcomp}
\end{figure}

In \hyperref[fig:agentcomp]{Fig.~\ref*{fig:agentcomp}}, we have box plots for each version of the agent that compare the times spent in the agent side of the algorithm.
This group of box plots follows the same structure as the previous one.
\hyperref[tab:agentcomp]{Table~\ref*{tab:agentcomp}} contains the average times.
We can see that even if only considering a DQN agent, there is a substantial difference between a version with online learning, and one without.
But, a Q-learning agent is faster even when it has to train and the DQN agent does not.

These results indicate that the agent choice has a significant impact on the execution speed.
However, these results are only informational, and are complemented by the final evaluation presented in \hyperref[4e]{Section~4.E}.

\subsection{Final Evaluation} \label{4e}

The previous tests were preliminary.
In this subsection, we compare our final DARA implementation with the simple approach.

Our final implementation used a DQN agent, with Python to read the files and parse the reward statistics file, and a Regex approach for parsing the state query file.
Additionally, it made use of the changes to the mac80211 module, which guarantees at least a 50~ms improvement, with room for improvement, if we consider a finer tuning.
This was compared to the simple DARA implementation, which did not consider the changes to the mac80211 module, and used the subprocess approach for the environment side.

\begin{table}[tbp]
    \caption{Average times of our simple and final DARA implementations.}
    \label{tab:finalcomp}
    \begin{center}
        \begin{tabular}{|c||c|c|c|}
            \hline
            Average times (ms) & Environment & Agent & Total  \\
            \hline \hline
            Simple Implementation & 578.770 & 20.670 & 599.440 \\
            \hline
            Final Implementation & 62.805 & 21.606 & 84.411 \\
            \hline
        \end{tabular}
    \end{center}
\end{table}

\hyperref[tab:finalcomp]{Table~\ref*{tab:finalcomp}} has the average times of an environment side step, an agent side step, and the total.
Note that there were no differences between agent side implementations, since our focus was on different computational implementations that do not change the design of the algorithm.
When contemplating and trying to reduce computational delays, DARA's environment side implementation is over nine times faster, and seven times faster overall.
We argue the larger delays in the simple DARA implementation are due to its frequent use of bash scripts, which tend to be slow, as well as the reliance in the subprocess module, which also introduces further overheads.
The simple implementation can take over half a second to react to changes in link quality.
This shows the relevance of implementations that contemplate the computational delays.

\section{Conclusions} \label{s5}

We identified the key operations present in state of the art RL-based RA algorithms, and chose DARA as an example application of our analysis methodology.
We presented an improved implementation of DARA that considers computational delays, and managed to reduce the execution time by one order of magnitude when compared to a simple implementation approach that disregards these delays.
This showed the importance of considering computational delays in this type of algorithms.
They are often overlooked in the state of the art.
By separating implementational details from their design, our improved implementation of DARA did not change its original concept, which indicates there is room for reducing computational delays without affecting the original design.
We confirmed that having lower level access to a device's data can remarkably reduce the delays of the algorithm.

Future work on this topic includes the application of the proposed analysis methodology to other RL-based RA algorithms and the measurement of the actual impact of the proposed changes in terms of network performance when running a specific RL-based RA algorithm.

\bibliographystyle{IEEEtran}
\bibliography{myrefs}

\end{document}